\documentclass[12pt]{iopart}

\usepackage{graphicx}
\usepackage[usenames,dvipsnames,svgnames,table]{xcolor}
\usepackage{float,subfigure}

\newcommand{\atanh  }{{\rm{atanh}}}
\newcommand{\atan   }{{\rm{arctan}}}

\newcommand{\bsigma}{\mbox{\boldmath$\sigma$}}
\newcommand{\btau}{\mbox{\boldmath$\tau$}}

\begin{document}

\title[Phase diagram and metastability of the Ising model on two coupled networks]{Phase diagram and metastability
  of the Ising model on two coupled networks}

\author{Ma\'ira Bolfe$^{1}$, Lucas Nicolao$^{2}$ and Fernando L. Metz$^{1,3,4}$}

\address{$^1$Departamento de F\'isica, Universidade Federal de Santa Maria, 97105-900 Santa Maria, Brazil}
\address{$^2$Departamento de F\'isica, Universidade Federal de Santa Catarina, 88040-900 Florian\'opolis,SC, Brazil}
\address{$^3$Instituto de F\'{\i}sica, Universidade Federal do Rio Grande do Sul, Caixa Postal 15051, 91501-970 Porto Alegre, Brazil}
\address{$^4$London Mathematical Laboratory, 14 Buckingham Street, London WC2N 6DF, United Kingdom}

\begin{abstract}
  We explore the cooperative behaviour and phase transitions 
  of interacting networks by studying
  a simplified model consisting of Ising spins placed on the nodes of two coupled
  Erd\"os-R\'enyi random graphs. We derive analytical
  expressions for the free-energy of the system and the magnetization of each graph, from which
  the phase diagrams, the stability of the different states, and the nature of the transitions among them, are clearly characterized. We show
  that a metastable state appears discontinuously by varying the model parameters, yielding 
  a region in the phase diagram where two solutions coexist. By performing Monte-Carlo
  simulations, we confirm the exactness of our main theoretical results and show that the typical time the system needs to escape
  from a metastable state grows exponentially fast as a function of the temperature, characterizing ergodicity breaking in the thermodynamic limit.
\end{abstract}

\vspace{2pc}
\noindent{\it Keywords}: Random graphs, complex networks, disordered systems

\section{Introduction}

Due to our increasing capability of collecting and manipulating large
amounts of data, it has become common sense that many real-world systems
are arranged in network structures \cite{BarratBook}.
Examples of problems defined on networks
are abundant in physics, biology, and finance, ranging from the inference
of ecological associations between microbial populations
\cite{Deng2012,Kurtz2015}
to the prediction of 
collapses in interbank networks \cite{Tiziano2013}. 

The study of phase transitions and critical phenomena in networked systems
constitutes an important research topic within the realm of complex networks \cite{Dorogo}.
Such phase transitions can be divided in two main
classes:
structural phase transitions, which refer to macroscopic changes in the architecture
features of networks, and phase transitions emerging due to the cooperative
behaviour of many entities interacting through the links of the network.
Examples of  structural phase transitions are the percolation
and condensation transitions \cite{Dorogo}, while synchronization \cite{Nishikawa,Gomez}
and formation of consensus  in social systems \cite{Castellano2009} are typical examples of cooperative phenomena.

Spin models of statistical physics are prototypical in the study
of the collective or cooperative behaviour of many interacting entities \cite{baxter}. Since in this case the individual
elements have a relatively simple mode of operation, one can usually deal with the intricate pattern
of  interconnections defining the network structure in a more detailed way \cite{Leone2002},
allowing to obtain qualitative conclusions
about the collective behaviour of the system
and the critical properties
characterizing eventual phase transitions. In particular, the Ising model \cite{baxter}, where each elementary unit
is represented by a binary variable,
has been used to characterize the formation of consensus in social systems \cite{Lambiotte2007}, opinion dynamics
and social spreading phenomena (see \cite{Castellano2009} and references therein).
In this context, each Ising spin represents an agent that is confronted with a binary decision
or choice \cite{Bouchaud2013}, which is taken based on the  choice of the majority in its local neighbourhood. The
graph connecting different spins reflects the social network structure, while the temperature mimics
the uncertainties of the agents or their idiosyncratic beliefs.

More recently, it has been realized that many real-world networks frequently do not operate in isolation, but
depend on the structure and dynamics of other networks \cite{Gao2012,Kivela2014}. 
This is the typical situation, for instance, in infrastructure networks, where the communication, electric power stations,
and transportation networks are coupled together  \cite{Gao2014}, in such a way that failure of nodes in one network
can lead to recursively disruption or malfunction of nodes on other networks, leading to a cascading of
failures \cite{Rosato2008,Buld2010}. Social networks are also commonly organized in modular or community
structures \cite{Girvan2002}, where the network is composed by sparsely coupled communities
or subgroups, with individuals densely connected inside each community. Networks of mobile phone
users \cite{Onnela2007} and of scientific collaborators \cite{Palla2005} are typical examples of social
networks with a modular structure.

The natural initial step to study the emergence of cooperative behaviour on coupled networks
is to consider models with Ising spins.
The effect of coupling two networks on the possible macroscopic states of the
system has been considered in
several works \cite{Suchecki2006,Dasgupta2009,Suchecki2009,Ostilli2009,Agliari2010,Chen2011,Huang2015}.
By performing Monte-Carlo simulations \cite{Dasgupta2009,Chen2011,Huang2015} and mean-field
approximations \cite{Suchecki2006,Suchecki2009}, it has been shown that
two possible solutions in terms of the magnetization of each network coexist: the
networks might be aligned (equal magnetizations) or anti-aligned (magnetizations with opposite signs), depending
on the model parameters. While the aligned state corresponds to the formation of consensus
  in a social system, the anti-aligned state
  represents the coexistence of contrary opinions between  two groups, which ultimately describes a polarized society.
  Since models with Ising spins give qualitative insights on
  real social systems, it is important to assess the robustness of the anti-aligned solution with respect to changes
  in the model parameters, to characterize the free-energy of the anti-aligned
  state in comparison to other solutions, and to study how its presence influences the dynamics of the system. 
  Apart from Monte-Carlo simulations suggesting the presence of metastable configurations on the pathway
  of coupled networks to the equilibrium \cite{Chen2011}, not much is known about any of the aforementioned aspects.
Phase diagrams illustrating the effect of the average connectivities on the macroscopic
behaviour are also absent from previous works \cite{Suchecki2006,Suchecki2009}.
These limitations stem from the naive mean-field approach employed in
references \cite{Suchecki2006,Suchecki2009}, which leads to a set of fixed-point equations for the magnetizations of each network, valid
strictly in the regime of large connectvities.

The aim of the present work is to fill this gap and fully explore the macroscopic behaviour of two interacting networks in contact
with a source of thermal noise.
We consider a  model
composed of Ising spins placed on the nodes of two interacting Erd\"os-R\'enyi random
graphs, with a ferromagnetic coupling between any pair of spins.
The model is simple enough to allow for a full analytical treatment and the results should
  yield valuable insights on the general behaviour of the Ising model on coupled networks. In other words, we expect
our work serves as a benchmark for studying more sophisticated models.

By using the replica approach of disordered systems, we derive the exact expressions
for the magnetizations of each network and for the free-energy of the system, which allows us to
obtain complete phase diagrams that unveil the role of the topology on the coexistence
  between ferromagnetic and anti-aligned states in this model. In particular, the
  region where both solutions coexist is strongly suppressed by an increase of the number
  of links among the two networks (see figure \ref{phasediagramT0}). We also show that the model displays a
  zero-temperature paramagnetic phase, essentially due
to the low average connectivity within each network and between them.
From the calculation of the free-energy
of the system, we show that the anti-aligned solution is always metastable
and it appears discontinuously as the model parameters are varied, clarifying
the stability properties of the macroscopic states in the coexistence region \cite{Suchecki2006,Suchecki2009}.
By means of Monte-Carlo simulations, we study the role of the metastability on the relaxation of the model to the equilibrium state by calculating
the average time $\tau$ the system needs to escape from a metastable initial state. The results for $\tau(T)$ as a function
of the temperature $T$ are described by the Vogel-Fulcher law $\ln \tau(T) \sim (T-T_0)^{-1}$, where the temperature
$T_0$ consistently converges, for increasing system size, to the instability temperature below which
metastable states are present in the thermodynamic limit. In the context of formation of consensus in social systems, our
  results indicate that, under
  certain conditions, two social groups can coexist with opposite opinions for remarkably long times, even if all
  interactions favour their agreement.
The exactness of our theoretical findings
is supported by Monte-Carlo simulations.

In the next section we define the model of coupled random graphs in equilibrium with a thermal bath. In section 
\ref{freeenergy} we present the main steps of the replica approach and the final
analytical expressions for the magnetizations of each network and the free-energy of the system.
The phase diagram and the stability of the macroscopic states are considered in section \ref{ssqw}, while
the results obtained through Monte-Carlo simulations are discussed in section \ref{secMC}. We
present some final remarks and perspectives in the last section.

\section{Random graph model of two coupled networks}

The model is composed of $2N$ interacting Ising spins or state variables. The spins $\sigma_{i} = \pm 1$ ($i=1,\dots,N$)
and $\tau_{i} = \pm 1$ ($i=1,\dots,N$) are coupled according to the following Hamiltonian
\begin{equation}
\mathcal{H}(\bsigma,\btau) = - J_{\sigma} \sum_{i < j}^{N}  c_{ij}^{\sigma} \sigma_i \sigma_j -  J_{\tau} \sum_{i < j}^{N} c_{ij}^{\tau} \tau_i \tau_j
- U  \sum_{i < j}^{N} c_{ij}^{I}  \left( \tau_i \sigma_j + \tau_j \sigma_i \right) \,,
\label{aaqp}
\end{equation}
where $\bsigma = (\sigma_1,\dots,\sigma_N)$ and  $\btau = (\tau_1,\dots,\tau_N)$.
The sum $\sum_{i < j}^{N} (\dots)$ runs over all distinct pairs of spins and the coupling
strengths ($J_{\sigma},J_{\tau},U$) are ferromagnetic.

The random variables $c_{ij}^{\sigma}$, $c_{ij}^{\tau}$ and $c_{ij}^{I}$ determine the topology of the model. 
We set $c_{ij}^{\sigma} = 1$ ($c_{ij}^{\tau} = 1$) if there is an edge between spins $\sigma_i$ ($\tau_i$) and $\sigma_j$ ($\tau_j$), and 
zero otherwise. The same definition applies to $c_{ij}^{I}$, which is responsible for the topology of connections among the two networks: we 
have $c_{ij}^{I} = 1$ if there is an edge between $\tau_i$ and $\sigma_j$ and between $\tau_j$ and $\sigma_i$, and 
$c_{ij}^{I} = 0$ otherwise. 
We consider the simplest network model, where these random variables are independently drawn from the distributions
\begin{eqnarray}
P_\sigma(c_{ij}^{\sigma}) = \prod_{i < j} \left[  \frac{c_{\sigma} }{N} \delta( c_{ij}^{\sigma},0)  + \left(1 -   \frac{c_{\sigma} }{N}  \right)  \delta( c_{ij}^{\sigma},1)  \right] \,, \label{hg1} \\
P_\tau(c_{ij}^{\tau}) = \prod_{i < j} \left[  \frac{c_{\tau} }{N} \delta(c_{ij}^{\tau},0)  + \left(1 -   \frac{c_{\tau} }{N}  \right)  \delta(c_{ij}^{\tau},1)  \right] \,, \label{hg2}  \\
P_I (c_{ij}^{I}) = \prod_{i < j} \left[  \frac{c_{I} }{N} \delta(c_{ij}^{I},0)  + \left(1 -   \frac{c_{I} }{N}  \right)  \delta(c_{ij}^{I},1)  \right] \,, \label{hg3} 
\end{eqnarray}
with $\delta$ representing the Kronecker delta. Essentially, the model is composed of two interacting Erd\"os-R\'enyi 
random graphs \cite{Bollobas}, where a given spin  interacts with a random subset of spins within 
its own graph and with a random subset of spins belonging to the second graph.
The parameter $c_{\sigma} > 0$ ($c_{\tau} > 0$) is the average number
of neighbours per node that belong to graph-$\sigma$ ($\tau$), while $c_{I}$ controls the average
number of edges per node connecting both graphs.  In
the limit $N \rightarrow \infty$, the number of edges per node within
each network, namely  $k^{(\sigma)}_{i} = \sum_{j =1 (\neq i)}^{N}  c_{ij}^{\sigma}$ and
$k^{(\tau)}_{i} = \sum_{j =1 (\neq i)}^{N}  c_{ij}^{\tau}$, follows a Poisson distribution
\begin{equation}
  p_{\sigma} (k) = \frac{c_{\sigma}^k  \exp{\left( -c_{\sigma}  \right)}  }{k!} \,,
  \qquad 
  p_{\tau} (k) = \frac{c_{\tau}^k  \exp{\left( -c_{\tau}  \right)}  }{k!} \,.  \label{pois1}
\end{equation}
The number of edges $k^{(I)}_{i} = \sum_{j =1 (\neq i)}^{N}  c_{ij}^{I}$ connecting a node $i$ in a certain
network to the nodes of the other network
also follows a Poisson distribution
\begin{equation}
  p_{I} (k) = \frac{c_{I}^k  \exp{\left( -c_{I}  \right)}  }{k!} \,. \label{pois2}
\end{equation}

The partition function of the system in equilibrium at temperature $T$ reads
\begin{equation}
\mathcal{Z} = \sum_{\bsigma, \btau} e^{- \beta \mathcal{H}(\bsigma,\btau)  } \,,
\end{equation}
with $\beta = T^{-1}$.
Our main objective consists in calculating the free-energy per spin in the thermodynamic limit $N \rightarrow \infty$.
Besides giving access to macroscopic observables, such as
the magnetization of each random graph, the free-energy allows us to clearly identify the
presence of metastable states. Assuming that, in the limit $N \rightarrow \infty$, the free-energy per spin $f$ is a 
self-averaging quantity with respect to fluctuations in the random graph structure, we have that
\begin{equation}
f = - \lim_{N \rightarrow \infty}   \frac{1}{2 \beta N} \langle  \ln \mathcal{Z}  \rangle \,,
\label{jj}
\end{equation}
in which $\langle \dots \rangle$ denotes the average over the ensemble of random graphs, defined
through eqs. (\ref{hg1}-\ref{hg3}).

%%%%%%%%%%%%%%%%%%%%%%%%%%%%%%%%%%%%%%%%%%%%%%%%%%%%%%%%%%%%%%%%%%%%%%%%%%%%%%%%%%%%%%%%%
%%%%%%%%%%%%%%%%%%%%%%%%%%%%%%%%%%%%%%%%%%%%%%%%%%%%%%%%%%%%%%%%%%%%%%%%%%%%%%%%%%%%%%%%%%

\section{The free-energy and the equations for the order-parameters} \label{freeenergy}

In order to calculate the ensemble average in eq. (\ref{jj}), we employ the replica method \cite{BookParisi}
\begin{equation}
\langle  \ln \mathcal{Z}  \rangle = \lim_{n \rightarrow 0} \frac{1}{n} \ln \langle \mathcal{Z}^{n}   \rangle\,.
\label{hhq}
\end{equation}
Initially, $n$ is considered to be an integer and positive exponent. After the ensemble average in eq. (\ref{hhq}) has
been evaluated and the limit $N \rightarrow \infty$ has been taken, the analytical
continuation $n \rightarrow 0$ yields
the free-energy per spin.
This is the standard strategy pursued in the replica approach \cite{BookParisi}. 
Since the microscopic states in this model are not frustrated, replica symmetry
yields exact results for the macroscopic behaviour of the system.
We remark that the cavity method \cite{MezPar,MezPar1}, also known as belief propagation in information
  theory \cite{MezardBook}, provides an alternative
tool to derive the same exact results as obtained in this section (see eqs. (\ref{free}), (\ref{field1}) and (\ref{field2})).

The calculation of the replicated partition function $\langle \mathcal{Z}^{n}   \rangle$ in the thermodynamic limit 
is analogous to previous models defined on random graphs \cite{Mezard87,Monasson,Leone2002}, so that we just present
here the main steps of the derivation. By computing the average over the random graph
ensemble in eq. (\ref{hhq}) and then introducing the order-parameters
\begin{eqnarray}
P_1(\bsigma) = \frac{1}{N} \sum_{i=1}^N \delta_{\bsigma,\bsigma_i} \,, \quad \bsigma = (\sigma_1,\dots,\sigma_n) \,, \label{P1} \\
P_2(\btau) = \frac{1}{N} \sum_{i=1}^N \delta_{\btau,\btau_i} \,, \quad \btau = (\tau_1,\dots,\tau_n) \,, \label{P2} \\
P_{12}(\bsigma,\btau) = \frac{1}{N} \sum_{i=1}^N \delta_{\btau,\btau_i} \delta_{\bsigma,\bsigma_i} \, \label{P12} ,
\end{eqnarray}
we are able to decouple sites in the expression for $\langle \mathcal{Z}^{n}   \rangle$, which can be recast in the integral form
\begin{eqnarray}
\fl
\langle   \mathcal{Z}^n  \rangle &\sim& \int \mathcal{D} \{ P,\hat{P} \}  \exp{\left(N g\left[  \{ P,\hat{P} \}   \right] \right)} \,,
\label{ggq}
\end{eqnarray}
with the integration measure 
\begin{equation}
\mathcal{D} \{ P,\hat{P} \} \equiv  \prod_{\bsigma \btau} dP_1(\bsigma) dP_2(\btau) dP_{12}(\bsigma,\btau) d\hat{P}_1(\bsigma) d\hat{P}_2(\btau) d\hat{P}_{12}(\bsigma,\btau)  \,.
\end{equation}
The functional $g\left[  \{ P,\hat{P} \}   \right]$ reads
\begin{eqnarray}
\fl
g\left[  \{ P,\hat{P} \}   \right] &=& - \frac{1}{2} \left(c_{\sigma} + c_{\tau} + c_{I}  \right) + i \sum_{\bsigma} P_1(\bsigma) \hat{P}_1(\bsigma) + i \sum_{\btau} P_2(\btau) \hat{P}_2(\btau)
+ i \sum_{\bsigma \btau} P_{12}(\bsigma,\btau) \hat{P}_{12}(\bsigma,\btau) \nonumber \\
\fl
&+& \frac{c_{\sigma}}{2} \sum_{\bsigma \bsigma^{\prime}} P_1(\bsigma) P_1(\bsigma^{\prime}) \exp{\left(\beta J_{\sigma} \bsigma.\bsigma^{\prime}\right)}
+ \frac{c_{\tau}}{2} \sum_{\btau \btau^{\prime}} P_2(\btau) P_2(\btau^{\prime}) \exp{\left( \beta J_{\tau}  \btau.\btau^{\prime}\right)} \nonumber \\
\fl
&+& \frac{c_{I}}{2} \sum_{\bsigma \bsigma^{\prime}}  \sum_{ \btau \btau^{\prime} } P_{12}(\bsigma,\btau)P_{12}(\bsigma^{\prime},\btau^{\prime}) 
\exp{\left[ \beta U \left(\bsigma.\btau^{\prime} + \bsigma^{\prime}.\btau  \right) \right]} \nonumber \\
\fl
&+& \ln{\left[ \sum_{\bsigma \btau}  e^{- i \hat{P}_1(\bsigma) - i \hat{P}_2(\btau) - i \hat{P}_{12}(\bsigma,\btau) }   \right]} \,,
\end{eqnarray}
where the conjugate parameters $\hat{P}_1$, $\hat{P}_2$ and $\hat{P}_{12}$ have arisen from the integral representations of
the Dirac delta functionals, used to introduce the order-parameters in the expression for
$\langle \mathcal{Z}^n \rangle$. From now on, the $n$-dimensional vector
$\bsigma$ ($\btau$) encodes the states of a single spin $\sigma_i$ ($\tau_i$) in the $n$ different replicas, as explicitly
emphasized in eqs. (\ref{P1}) and (\ref{P2}). Unimportant factors, which give a vanishing contribution to
the free-energy per spin in the limit $N \rightarrow \infty$, have been neglected in eq. (\ref{ggq}).

The function $\langle \mathcal{Z}^n \rangle$ can now be evaluated through the saddle-point method. In
the limit $N \rightarrow \infty$, the integral in eq. (\ref{ggq}) is dominated by 
the values of $\{ P,\hat{P} \}$ that extremize the functional $g \left[  \{ P,\hat{P} \} \right]$.
Substituting eq. (\ref{hhq}) in eq. (\ref{jj}) and performing the limit
$N \rightarrow \infty$ through the saddle-point method, we obtain a formal expression for
the free-energy per spin
\begin{equation}
  2 \beta f = - \lim_{n \rightarrow 0} \frac{1}{n} g\left[  \{ P,\hat{P} \} \right] \,, \label{gaa}
\end{equation}  
where $\{ P,\hat{P} \}$ refers, from now on, to the specific values that extremize $g\left[  \{ P,\hat{P} \} \right]$. The
saddle-point equations
that determine $\{ P,\hat{P} \}$ are derived by taking functional derivatives of $g\left[  \{ P,\hat{P} \} \right]$ with respect
to $\{ P,\hat{P} \}$
\begin{eqnarray}
P_{1}(\bsigma) = \frac{1}{\mathcal{N}} \sum_{\btau} \exp{\left[ -i \hat{P}_{1}(\bsigma) -i \hat{P}_{2}(\btau) - i  \hat{P}_{12}(\bsigma,\btau) \right] }  \, , \label{kjh1} \\
P_{2}(\btau) = \frac{1}{\mathcal{N}} \sum_{\bsigma} \exp{\left[ -i \hat{P}_{1}(\bsigma) -i \hat{P}_{2}(\btau) - i  \hat{P}_{12}(\bsigma,\btau) \right] } \, , \label{kjh2} \\
P_{12}(\bsigma,\btau) = \frac{1}{\mathcal{N}} \exp{\left[ -i \hat{P}_{1}(\bsigma) -i \hat{P}_{2}(\btau) - i  \hat{P}_{12}(\bsigma,\btau) \right] } \,, \label{kjh3}
\end{eqnarray}
where $\mathcal{N}$ is the normalization factor
\begin{equation}
\mathcal{N} = \sum_{\bsigma \btau}\exp{\left[ -i \hat{P}_{1}(\bsigma) -i \hat{P}_{2}(\btau) - i  \hat{P}_{12}(\bsigma,\btau) \right] } \,.
\end{equation}
The conjugate parameters are given by
\begin{eqnarray}
  \hat{P}_1(\bsigma) = i c_{\sigma} \sum_{\bsigma^{\prime}} P_1(\bsigma^{\prime}) \exp{\left( \beta J_{\sigma}
    \bsigma^{\prime}.\bsigma \right)} \,, \label{tt1} \\
\hat{P}_2(\btau) = i c_{\tau} \sum_{\btau^{\prime}} P_2(\btau^{\prime}) \exp{\left( \beta J_{\tau} \btau^{\prime}.\btau \right)} \,, \label{tt2} \\
\hat{P}_{12}(\bsigma,\btau) = i c_{I} \sum_{\bsigma^{\prime} \btau^{\prime}} P_{12}(\bsigma^{\prime},\btau^{\prime}) \exp{\left[  \beta U 
\left( \bsigma.\btau^{\prime} + \bsigma^{\prime}.\btau \right)  \right]} \,. \label{tt3}
\end{eqnarray}
From eqs. (\ref{gaa}-\ref{tt3}), we see that the free-energy per spin is fully determined by the self-consistent
equations (\ref{kjh1}-\ref{kjh3}) for the order-parameters.

In order to compute the limit $n \rightarrow 0$ in eq. (\ref{gaa}), one has to explicitly perform the sums over the replica 
Ising spins and unveil how $g\left[  \{ P,\hat{P} \} \right]$ depends on $n$, which is only possible if we
make an assumption for the structure of the order-parameters. By considering there is one
single thermodynamic state, each order-parameter is invariant with respect to permutations  
of the replica indexes \cite{Monasson} and all information about the fluctuations of
the local magnetizations lies in the distribution of effective
fields $h_i = \beta^{-1} \atan{\left( \left\langle S_i   \right\rangle \right)}$, where $\langle S_i \rangle$
denotes the average of a generic spin $S_i$ with respect to thermal
and random graph fluctuations.
The simplest form  that
fulfills replica symmetry is a function of the magnetizations only, namely \cite{Monasson,Leone2002}
\begin{eqnarray}
P_1(\bsigma) = \int dh \, W_{\sigma}(h) \, \frac{\exp{\left(  \beta h \sum_{\alpha=1}^n \sigma_\alpha \right)}}{\left[ 2 \cosh{ \left( \beta h  \right) }  \right]^n   } \,,  \label{mm1} \\
P_2(\btau) = \int dh \, W_{\tau}(h) \, \frac{\exp{\left(  \beta h \sum_{\alpha=1}^n \tau_\alpha \right)}}{\left[ 2 \cosh{ \left( \beta h  \right) }  \right]^n   } \, ,  \label{mm2} \\
P_{12}(\bsigma,\btau) = \int du dv \, W_{\sigma \tau}(u,v) \, \frac{\exp{\left(  \beta u \sum_{\alpha=1}^n \sigma_\alpha + \beta v \sum_{\alpha=1}^n \tau_\alpha  \right)}}
{\left[ 4 \cosh{ \left( \beta u  \right) }   \cosh{ \left( \beta v  \right) }  \right]^n   } \,.  \label{mm3} 
\end{eqnarray}
The quantity $W_{\sigma}(h)$ ($W_{\tau}(h)$) is the distribution of effective fields
on network-$\bsigma$ ($\btau$), independently 
of the configuration of effective fields in network-$\btau$ ($\bsigma$).
The distributions $W_{\sigma}(h)$ and $W_{\tau}(h)$
are normalized, consistently with eqs. (\ref{P1}) and (\ref{P2}). 
The function $W_{\sigma \tau}(u,v)$ is the joint distribution of effective fields in both networks, where the argument $u$ ($v$)
refers to the possible outcomes for the effective fields 
in network-$\bsigma$ ($\btau$). Besides the normalization
of $W_{\sigma \tau}(u,v)$, we have to supplement 
eq. (\ref{mm3}) with the conditions $\int du \, W_{\sigma \tau}(u,v) = W_\tau (v)$ and $\int dv \, W_{\sigma \tau}(u,v) = W_\sigma (u)$, which
ensure the marginalization of $P_{\sigma \tau}(\bsigma,\btau)$
with respect to the spins of a given network, consistently with eqs. (\ref{P1}-\ref{P12}).

By substituting eqs. (\ref{mm1}-\ref{mm3}) in eq. (\ref{gaa}), computing the trace
over the Ising spins, and performing
the limit $n \rightarrow 0$, we obtain the free-energy per spin
\begin{eqnarray}
  \fl
  f &=&  c_{\sigma}  \int d h d h^{\prime} W_{\sigma}(h) W_{\sigma}(h^{\prime})
  \mathcal{U}_{\beta}(h,h^{\prime}|J_{\sigma})
+  c_{\tau}  \int d h d h^{\prime} W_{\tau}(h) W_{\tau}(h^{\prime})
\mathcal{U}_{\beta}(h,h^{\prime}|J_{\tau})  \nonumber \\
\fl
&+&  2 c_{I}  \int d h d h^{\prime} W_{\tau}(h) W_{\sigma}(h^{\prime})
  \mathcal{U}_{\beta}(h,h^{\prime}|U)
  - \frac{1}{2 \beta} \sum_{k_{\sigma},k_{I}=0}^{\infty}  p_{\sigma}(k_{\sigma}) p_{I}(k_{I}) \nonumber \\
  \fl
  &\times&
  \int \left( \prod_{n=1}^{k_{\sigma}} d u_n W_{\sigma}(u_n)   \right) \left( \prod_{m=1}^{k_{I}} d v_m W_{\tau}(v_m)   \right)
\ln{\left[ \sum_{\gamma= \pm 1}
\mathcal{G}_{\gamma}(u_{1},\dots,u_{k_{\sigma}}|J_{\sigma}) \mathcal{G}_{\gamma}(v_{1},\dots,v_{k_{I}}|U)
    \right]} \nonumber \\
\fl
&-&
\frac{1}{2 \beta} \sum_{k_{\tau},k_{I}=0}^{\infty}  p_{\tau}(k_{\tau}) p_{I}(k_{I})
\int \left( \prod_{n=1}^{k_{\tau}} d u_n W_{\tau}(u_n)   \right) \left( \prod_{m=1}^{k_{I}} d v_m W_{\sigma}(v_m)   \right) \nonumber \\
\fl
&\times& \ln{\left[ \sum_{\gamma= \pm 1}   
\mathcal{G}_{\gamma}(u_{1},\dots,u_{k_{\tau}}|J_{\tau}) \mathcal{G}_{\gamma}(v_{1},\dots,v_{k_{I}}|U)
    \right]}\,, \label{free}
\end{eqnarray}
where
\begin{eqnarray}
  \fl
\mathcal{U}_{\beta}(u,v|J) = \frac{1 }{ 4 \beta }
\ln{\left[ \frac{  1 + \tanh{\left( \beta u \right)} \tanh{\left( \beta v \right)} \tanh{\left( \beta J \right)}   }
    {  \cosh{\left( \beta J \right)}  }  \right]}\,, \nonumber \\
\fl
\mathcal{G}_{\gamma}(u_{1},\dots,u_{K}|J) =
  \prod_{n=1}^{K}  
    \Big[ 1 + \gamma \tanh\left( \beta u_n   \right) \tanh\left( \beta J    \right)    \Big]\,,  \nonumber
\end{eqnarray}
with $\gamma \in \{ -1,1 \}$. Note that $f$ is independent of the joint distribution of effective
fields  $W_{\sigma \tau}(u,v)$, so that it suffices to derive self-consistent equations for the
distributions  $W_{\sigma}(h)$ and  $W_{\tau}(h)$. These are obtained by
plugging eqs. (\ref{mm1}-\ref{mm3}) in eqs. (\ref{kjh1}-\ref{kjh2}) and taking
the limit $n \rightarrow 0$
\begin{eqnarray}
  \fl
  W_{\sigma} (h) &=& \sum_{k_{\sigma},k_{I} =0}^{\infty} p_{\sigma}(k_{\sigma}) p_{I}(k_{I})
  \int \left( \prod_{n=1}^{k_{\sigma}} d h_n W_{\sigma}(h_n)   \right)\left( \prod_{m=1}^{k_{I}} d h_m W_{\tau}(h_m)   \right) \nonumber \\
  \fl
  &\times&
  \delta \Big[ h - \mathcal{F}_{\beta} \left(h_{1},\dots, h_{k_{\sigma}} | J_{\sigma}    \right)
    - \mathcal{F}_{\beta} \left(h_{1},\dots, h_{k_{I}} | U  \right) \Big] \,, \label{field1}
\end{eqnarray}
  \begin{eqnarray}
    \fl
    W_{\tau} (h)& =& \sum_{k_{\tau},k_{I} =0}^{\infty} p_{\tau}(k_{\tau}) p_{I}(k_{I})
  \int \left( \prod_{n=1}^{k_{\tau}} d h_n W_{\tau}(h_n)   \right)\left( \prod_{m=1}^{k_{I}} d h_m W_{\sigma}(h_m)   \right) \nonumber \\
  \fl
  &\times&
  \delta\Big[ h - \mathcal{F}_{\beta} \left(h_{1},\dots, h_{k_{\tau}} | J_{\tau}    \right)
    - \mathcal{F}_{\beta} \left(h_{1},\dots, h_{k_{I}} | U  \right) \Big] \,, \label{field2}  
  \end{eqnarray}
where the degree distributions $p_{\sigma}(k)$, $p_{\tau}(k)$ and $p_{I}(k)$ are defined
in eqs. (\ref{pois1}-\ref{pois2}), while the function $\mathcal{F}_{\beta}$ reads
\begin{equation}
  \mathcal{F}_{\beta} \left(h_{1},\dots, h_{K} | J    \right) =
  \frac{1}{\beta}  \sum_{i=1}^{K}  \atanh\Big( \tanh{\left(  \beta h_i \right)}
 \tanh{\left(  \beta J \right)}   \Big) \,.
\end{equation}
The magnetizations of each network are given by
\begin{equation}
  m_{\sigma} = \frac{1}{N} \sum_{i=1}^N \langle \sigma_i  \rangle
  \qquad m_{\tau} = \frac{1}{N} \sum_{i=1}^N \langle \tau_i  \rangle \,.  \nonumber 
\end{equation}
In the present formalism, the magnetizations are obtained from the effective field distributions \cite{Monasson}:
\begin{eqnarray}
  m_{\sigma} = \int dh W_{\sigma}(h) \tanh(\beta h)\,, \nonumber \\
   m_{\tau} = \int dh W_{\tau}(h) \tanh(\beta h)\,. \nonumber
\end{eqnarray}
Thus, once  $W_{\sigma}(h)$ and  $W_{\tau}(h)$ are determined from the solutions of
eqs. (\ref{field1}) and (\ref{field2}), we can calculate the magnetizations of each
network, obtain the phase diagrams, and probe the stability of the
solutions through the free-energy.

\section{Phase diagrams and metastability} \label{ssqw}

For a general combination of model parameters, eqs. (\ref{field1}) and (\ref{field2}) cannot be analytically solved and one needs to employ
a numerical approach. In this section we solve numerically eqs. (\ref{field1}) and (\ref{field2}) through
the population dynamics method \cite{MezPar,MezPar1}, from which the phase diagrams and the free-energy follow.
In this numerical approach, the distributions $W_{\sigma}(h)$ and  $W_{\tau}(h)$
are parametrized, respectively, by large sets of stochastic
variables $\{ h_{i}^{(\sigma)}  \}_{i=1,\dots,\mathcal{N}}$ and $\{ h_{i}^{(\tau)}  \}_{i=1,\dots,\mathcal{N}}$, with
$\mathcal{N}$ denoting the population size.
By choosing initial
distributions  $W^{(0)}_{\sigma}(h)$ and  $W^{(0)}_{\tau}(h)$ for each network, the variables
 $\{ h_{i}^{(\sigma)}  \}_{i=1,\dots,\mathcal{N}}$ and $\{ h_{i}^{(\tau)}  \}_{i=1,\dots,\mathcal{N}}$
are consistently updated according to the arguments of the Dirac delta functions appearing
in eqs. (\ref{field1}) and (\ref{field2}), until the empirical distribution obtained from each population
of fields reaches its final, stationary form. Averages
involving $W_{\sigma}(h)$ and $W_{\tau}(h)$ are evaluated by computing sample averages
using, respectively, the collection of random variables $\{ h_{i}^{(\sigma)}  \}_{i=1,\dots,\mathcal{N}}$ and $\{ h_{i}^{(\tau)}  \}_{i=1,\dots,\mathcal{N}}$.
We refer to \cite{MezardBook} for further details regarding this numerical method.

From the numerical solutions of eqs. (\ref{field1}) and (\ref{field2}), we have calculated
the magnetizations $m_{\sigma}$ and $m_{\tau}$ of each network for different values of the
model parameters. Three different solutions have been found: a paramagnetic state (P), with
$m_{\sigma}=m_{\tau} =0$; a ferromagnetic solution (F), where $m_{\sigma}m_{\tau} > 0$; and
an anti-aligned state (AA), with $m_{\sigma}m_{\tau} < 0$.
\begin{figure}
\begin{center}
  \includegraphics[scale=0.5]{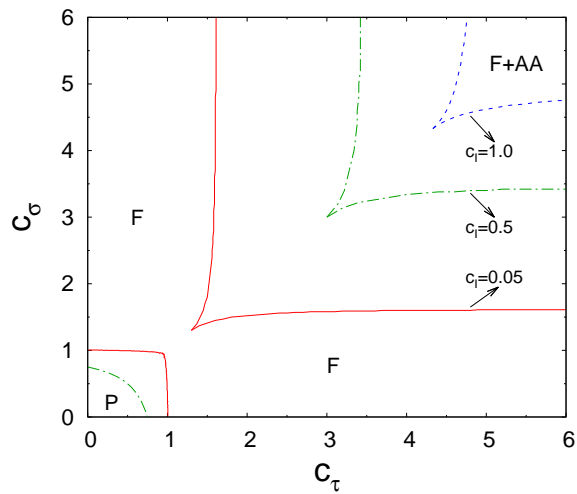}
  \caption{Phase diagram in the plane $(c_{\sigma},c_{\tau})$ for temperature $T=0.001$, a value $U=0.1$ for the coupling
    strength between the networks,
    and different values of the average connectivity $c_I$ between the networks. The model displays
    a ferromagnetic solution (F), an anti-aligned (AA) solution, and a paramagnetic (P) state.
    The ferromagnetic and the anti-aligned solutions coexist in the region
    marked with F+AA, where the AA solution is always metastable.  
  These results are obtained through the numerical solution of eqs. (\ref{field1}) and (\ref{field2})
  using the population dynamics method with $\mathcal{N} = 5 \times 10^5$ and initial distributions
  $W^{(0)}_{\sigma}(h) = \delta(h-1)$ and   $W^{(0)}_{\tau}(h) = \delta(h+1)$ (see the main text).
}
\label{phasediagramT0}
\end{center}
\end{figure}

In order to discuss the phase diagrams and the stability of these macroscopic states, we set $J_\sigma = J_\tau = 1$ throughout this section.
Figure \ref{phasediagramT0} shows typical phase diagrams in the $(c_{\sigma}, c_{\tau})$-plane for low temperatures and
different values of $c_I$. The networks are weakly coupled with strength $U = 0.1$.
For $J_\sigma = J_\tau$, the Hamiltonian is invariant
with respect to the interchange of the adjacency matrix elements $c_{ij}^{\sigma} \leftrightarrow c_{ij}^{\tau} \,\, \forall \, i \,, j$, which
implies on the symmetry of the above phase diagram around the straight line $c_{\sigma} = c_{\tau}$.
In the region F+AA of
figure \ref{phasediagramT0}, the ferromagnetic
solution coexists with the (metastable) anti-aligned state \cite{Suchecki2006,Suchecki2009}, namely, both
types of order are obtained from the numerical solution of eqs. (\ref{field1}) and (\ref{field2}), depending
on the initial distributions  $W^{(0)}_{\sigma}(h)$ and  $W^{(0)}_{\tau}(h)$ in the population dynamics method.
As shown in figure \ref{phasediagramT0}, the anti-aligned solution is more robust or more abundant in
the phase diagram for strong modularity, i.e., when $c_I$ is much smaller than the mean connectivities
within each network. In the paramagnetic phase, the effective field distributions are given
by  $W_{\sigma}(h) = W_{\tau}(h) = \delta(h)$. The distributions $W_{\sigma}(h)$ and $W_{\tau}(h)$ have
a small average close to the boundary between the P and F phases, which allows to expand the right
hand side of eqs. (\ref{field1}) and (\ref{field2}) and derive the equation for the boundary
between these states
\begin{equation}
  c_I^2 \tanh^2{\left( \beta U  \right)} = \left[ 1 - c_{\sigma}  \tanh{\left( \beta J_{\sigma}  \right)}    \right]
  \left[ 1 - c_{\tau}  \tanh{\left( \beta J_{\tau}  \right)}    \right]\,.
  \label{fpq}
\end{equation}
From the above equation, one concludes that the system exhibits a paramagnetic phase
for $T=0$, located in a region of the phase diagram where all average connectivities must be smaller than
one. This is consistent with figure \ref{phasediagramT0}, in which the P phase is absent for $c_I = 1$. The
reason for the existence of this zero-temperature P phase is utterly topological, since for low
connectivities the system is fragmented in a large number of finite non-interacting clusters \cite{Dorogo}.

\begin{figure}[H]
\begin{center}
  \includegraphics[scale=0.5]{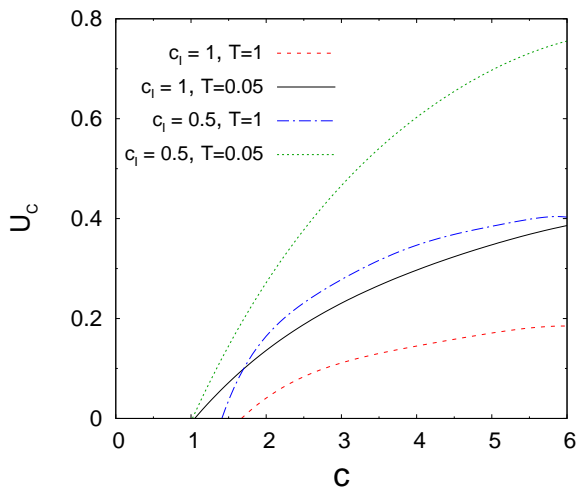}
  \caption{Critical coupling strength $U_c$ below which we find an anti-aligned solution
from eqs. (\ref{field1}) and (\ref{field2}). The results
    are shown as a function of the average connectivity $c \equiv c_{\sigma} = c_{\tau}$ within each network, for different combinations
    of temperature $T$ and the average connectivity $c_I$ between the networks. We have rescaled all coupling
    constants (see eq. (\ref{aaqp})) by the common factor $c_{\rm eff} = \frac{1}{3} (c_{\sigma} + c_{\tau} + c_I )$. 
     These results are obtained through the numerical solution of eqs. (\ref{field1}) and (\ref{field2})
     using the population dynamics method with $\mathcal{N} = 5 \times 10^5$ and initial distributions
   $W^{(0)}_{\sigma}(h) = \delta(h-1)$ and   $W^{(0)}_{\tau}(h) = \delta(h+1)$ (see the main text).
}
\label{Uversusc}
\end{center}
\end{figure}

Figure \ref{Uversusc} complements the phase diagram of figure \ref{phasediagramT0} by showing
results for the critical coupling strength $U_c$ above which the anti-aligned solution is absent, considering
different values of $T$ and $c_I$. The
parameter $c \equiv c_\tau = c_\sigma$ is the average connectivity within each network.
As can be noted, the increase of $T$ or $c_I$ has a detrimental effect on the existence of anti-aligned states.
Below we study
in more detail the effect of thermal fluctuations in the stability of such states.
The curves in figure \ref{Uversusc} converge to $c=c_{\rm perc}(T)$ as $U \rightarrow 0$, where
$c_{\rm perc}(T)$ is the critical average connectivity above which a single random graph lies
in a ferromagnetic state.
We have that $c_{\rm perc}(T) \simeq 1$ for the smallest temperature displayed in  figure \ref{Uversusc}, consistent
with standard results for the percolation transition in random graphs  \cite{Dorogo}.

\begin{figure}[H]
\begin{center}
\subfigure[Magnetization.]{
  \includegraphics[width=0.45\textwidth]{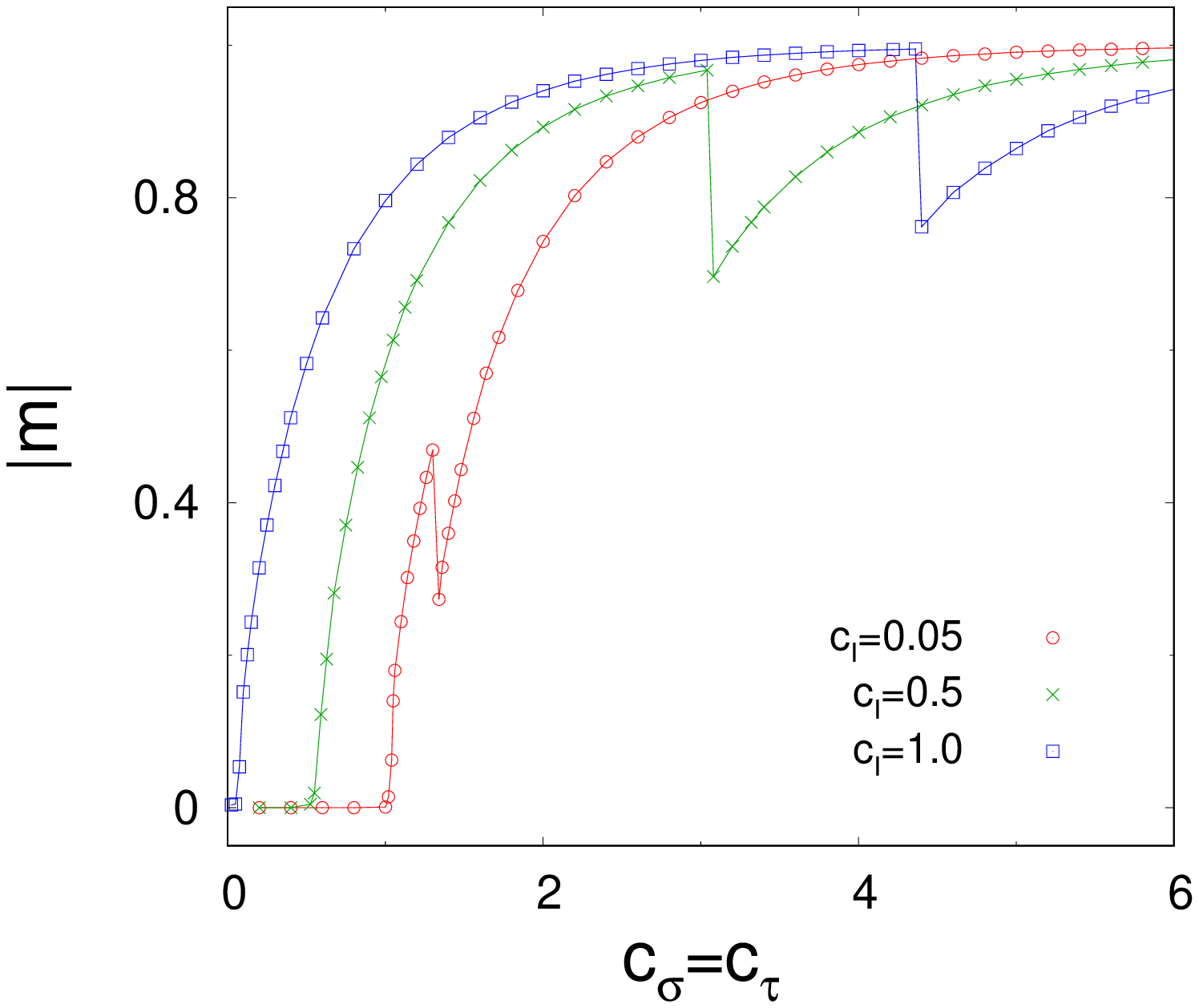}
\label{Mt0}
}
  \subfigure[Free-energy for $c_I=0.5$.]{
    \includegraphics[width=0.45\textwidth]{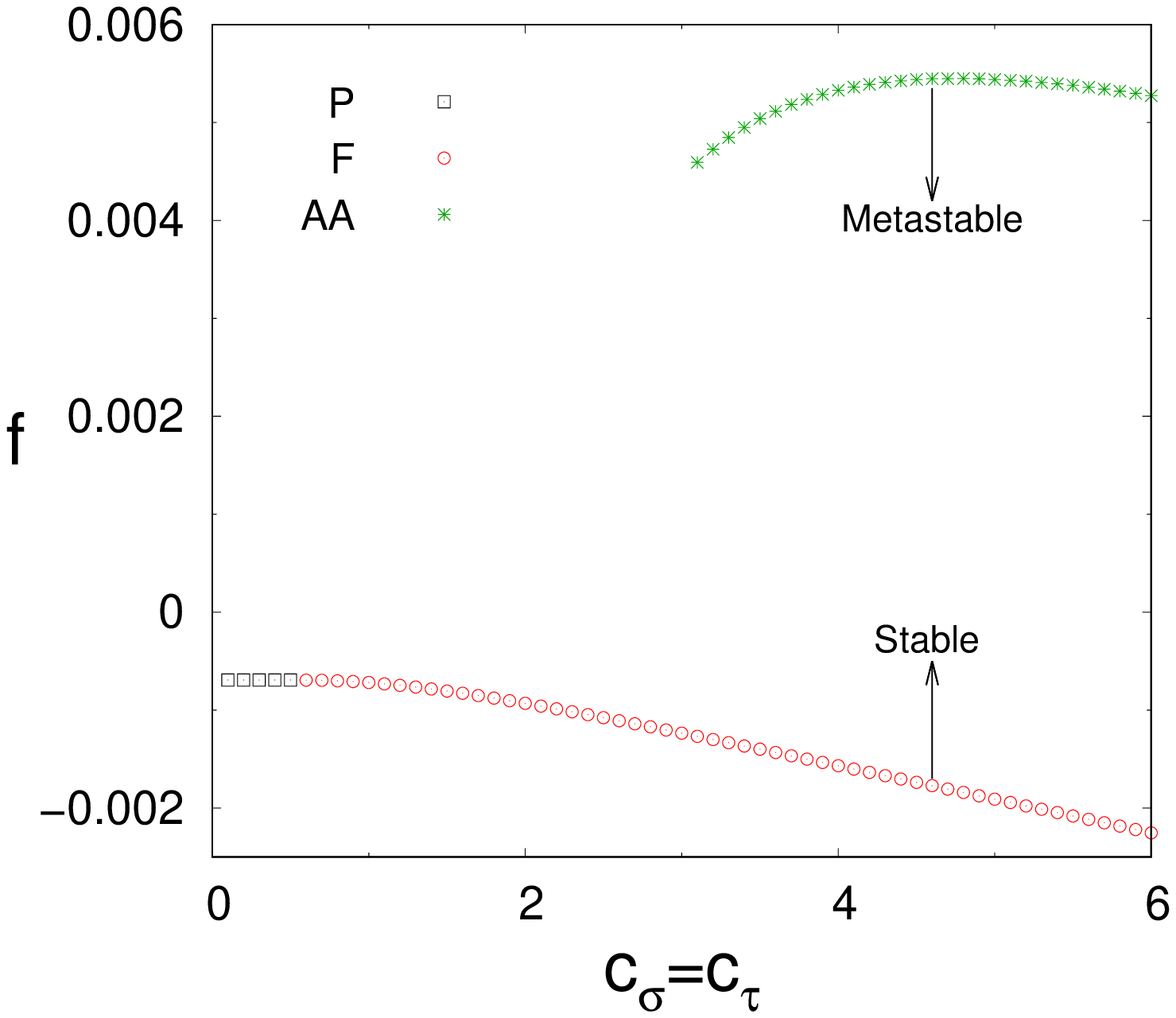}
\label{freet0}
  }
  \caption{Absolute value of the magnetization $|m|$ of a single network and free-energy per spin $f$ as functions of
    $c_{\sigma} = c_{\tau}$ for temperature $T=0.001$ and ferromagnetic coupling $U=0.1$. The values
    of the average connectivity $c_I$ between the networks are indicated on the graphs.
    The magnetizations of each network have the same absolute value for $c_{\sigma} = c_{\tau}$. The three
    different solutions of the phase diagram and the corresponding free-energies are shown here. These results are obtained through
  the population dynamics method with $\mathcal{N} = 10^6$  and initial distributions
  $W^{(0)}_{\sigma}(h) = \delta(h-1)$ and   $W^{(0)}_{\tau}(h) = \delta(h+1)$ (see the main text).
  }
\label{MagT0}
\end{center}
\end{figure}

In figure \ref{Mt0} we show the absolute value of the magnetization of a single network
along the straight line $c_{\sigma} = c_{\tau}$ of the phase diagram, considering
initial distributions $W^{(0)}_{\sigma}(h) = \delta(h-1)$ and   $W^{(0)}_{\tau}(h) = \delta(h+1)$
that yield $m_{\sigma} >0$ and $m_{\tau} <0$ in the region F+AA.
As clearly shown, the transition between the paramagnetic and
the F state is continuous, while the magnetization changes discontinuously along the boundary between the F region and the F+AA region. In this 
case, such discontinuity is not a signature of a first-order phase transition, but it
simply reflects the sudden emergence of the metastable anti-aligned solution. The equilibrium magnetization, characterized
by a continuous branch, is not shown when $c_{\sigma} = c_{\tau}$ lies in the F+AA region, since eqs. (\ref{field1}) and (\ref{field2}) have been
solved with initial distributions favouring the anti-aligned solution. The stability of the different
solutions is characterized in figure \ref{freet0}, where we present the free-energy $f$ of each possible
solution of eqs. (\ref{field1}) and (\ref{field2}) as a function of  $c_{\sigma} = c_{\tau}$, for a
single value of $c_{I}$ \footnote{The free-energy of the paramagnetic solution is displayed only
for model parameters within the region P, where the paramagnetic state is the only possible solution
of eqs.  (\ref{field1}) and (\ref{field2}). In regions F and F+AA, figure \ref{freet0} exhibits
only the free-energy of the nontrivial states.}.
The main outcome is that the anti-aligned solution is always metastable, while the ferromagnetic
solution is the stable macroscopic state, since it corresponds to the global minimum of the free-energy.
We have checked many different combinations
of model parameters and we did not find any qualitative changes in these stability properties.
All results discussed in this section are also applied to the case where the couplings between
the networks are anti-ferromagnetic. The difference is that, for $U < 0$, the ferromagnetic solution
is metastable, while the anti-aligned solution is the stable macroscopic state. Apart
from that, the phase diagrams remain unchanged, i.e., the phase boundaries for $U < 0$ are the same as those for $U > 0$.

\section{Numerical simulations} \label{secMC}

In this section we compare our theoretical results with Monte-Carlo
(MC) simulations using standard Metropolis dynamics of finite size
systems with different system sizes. We also compute the typical time
the system needs to escape from the metastable states. This is a way to
go beyond the theoretical results and quantify the lifetime of the
metastable states and have a better idea of their role on the
relaxation of the system to equilibrium.

The theoretical results indicating a second order phase transition
from a paramagnetic to a ferromagnetic phase, where metastable
anti-aligned states appear for certain model parameters, are verified
in the MC simulations. We have measured the
magnetizations of each network in equilibrium or in metastable
configurations, following a quasi-static heating protocol of the system
prepared initially at zero temperature in a purely anti-aligned metastable state,
where the spins in different networks have opposite directions.
Simulations were done for
$J_{\sigma} = J_{\tau} = U=1/c_{\rm eff}$, with $c_{\rm eff} = \frac{1}{3}(c_{\sigma} + c_{\tau} + c_I )$, and two cases of average
connectivities $c_{\sigma}=c_{\tau}=10$ and $c_I = 1.0$, and
$c_{\sigma}=c_{\tau}=4$ and $c_I = 0.5$.  Simulated system sizes range
from $N=400$ up to $N=25600$, where $N$ stands for the number of nodes
in each graph.  Equilibration times in each temperature are of order
10$^5$ MC steps, and averages are taken from 100 different
realizations of the random graphs, each realization contributing with
100 samples for each temperature.

\begin{figure}[H]
\begin{center}
\subfigure[$\,$ $c_{\sigma}=c_{\tau}=10$ and $c_I = 1.0$.]{
  \includegraphics[width=0.475\textwidth]{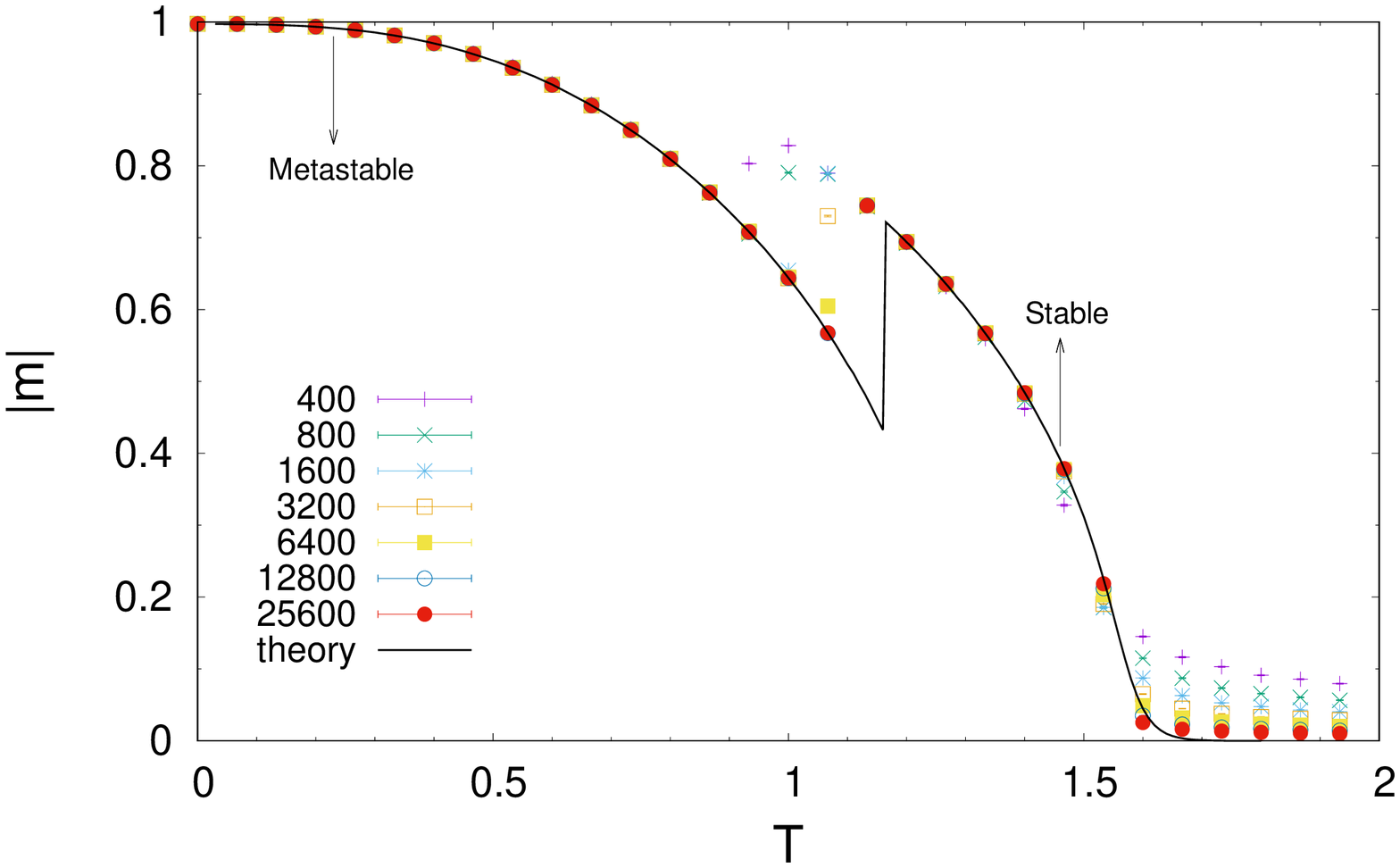}
\label{ssd1}
}
  \subfigure[$\,$ $c_{\sigma}=c_{\tau}=4$ and $c_I = 0.5$.]{
    \includegraphics[width=0.475\textwidth]{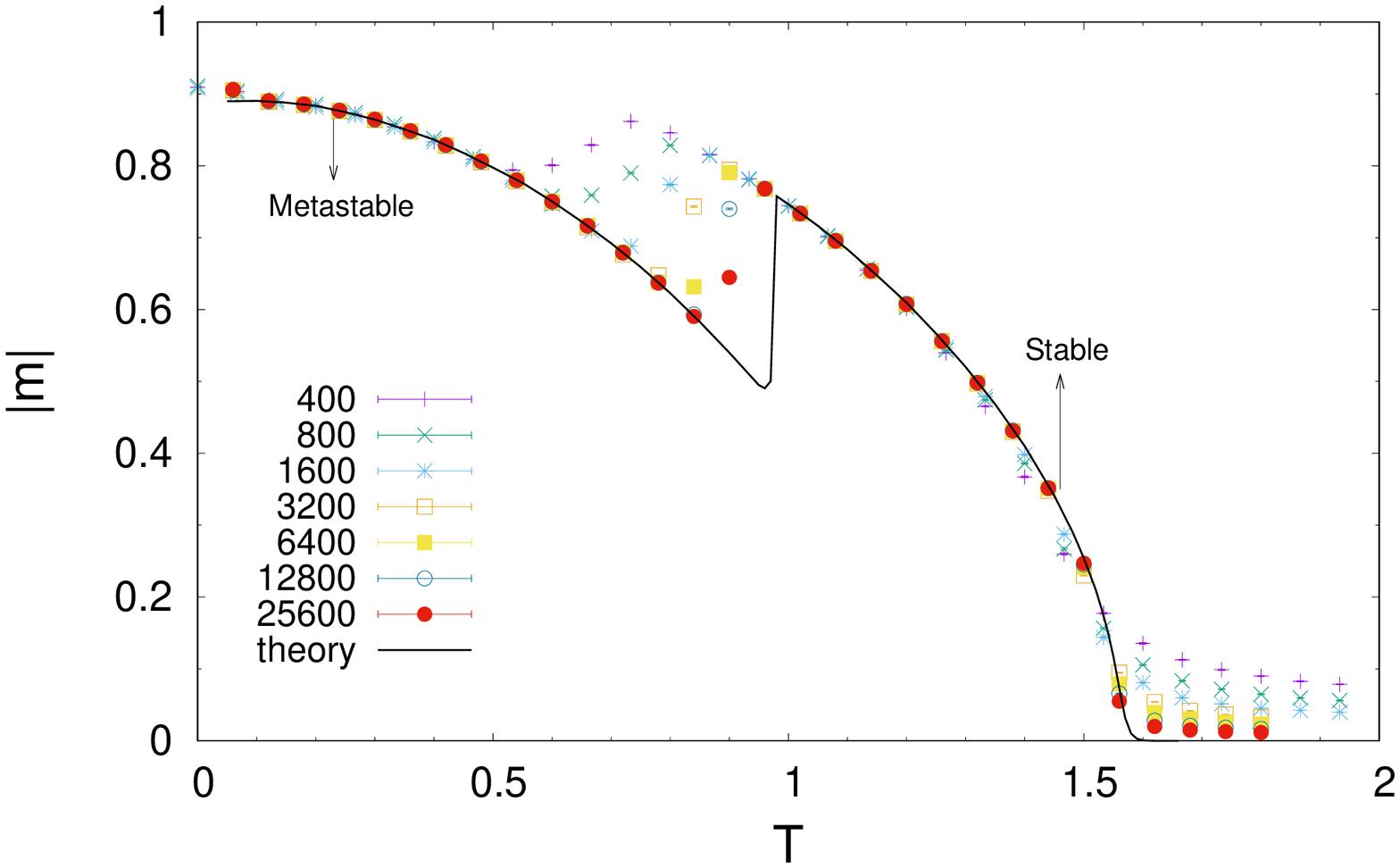}
\label{ssd2}
  }
  \caption{Results for the magnetization $|m|$ of each network as a
    function of the temperature $T$ for coupling strengths $J_{\sigma}
    = J_{\tau} = U=1/c_{\rm eff}$, with $c_{\rm eff} =
    \frac{1}{3}(c_{\sigma} + c_{\tau} + c_I )$. The magnetization of
    each graph has the same absolute value for
    $c_{\sigma}=c_{\tau}$. Each panel compares the theoretical
    results, derived from the solutions of eqs. (\ref{field1}) and
    (\ref{field2}), with Monte-Carlo simulations following a heating
    protocol started at $T=0$ in the metastable anti-aligned state, 
    for different systems sizes $N$, whose values are shown on
    each graph. Equations (\ref{field1}) and (\ref{field2}) are solved
    numerically using the population dynamics method with $\mathcal{N} = 10^6$ and
    initial distributions $W^{(0)}_{\sigma}(h) = \delta(h-1)$ and   $W^{(0)}_{\tau}(h) = \delta(h+1)$  (see the main text).     
  }
\label{MC}
\end{center}
\end{figure}

Figure \ref{MC} shows the comparison between MC simulations and our
theoretical results for the magnetization of each network as a
function of the temperature. As can be noticed, finite size effects in
MC simulations become remarkable as $T$ increases towards the
instability temperature, above which the anti-aligned metastable
  state disappears. This is a purely dynamical effect in the heating
protocol, due to the available thermal energy and finite energy
barrier between the metastable state and the true thermodynamical
equilibrium state. In spite of that, the overall agreement between
our theory and MC simulations is excellent, with the simulation data
consistently approaching the theoretical curves for increasing $N$.

\begin{figure}[H]
\begin{center}
  \subfigure[]{
  \includegraphics[width=0.475\textwidth,height=0.32\textwidth]{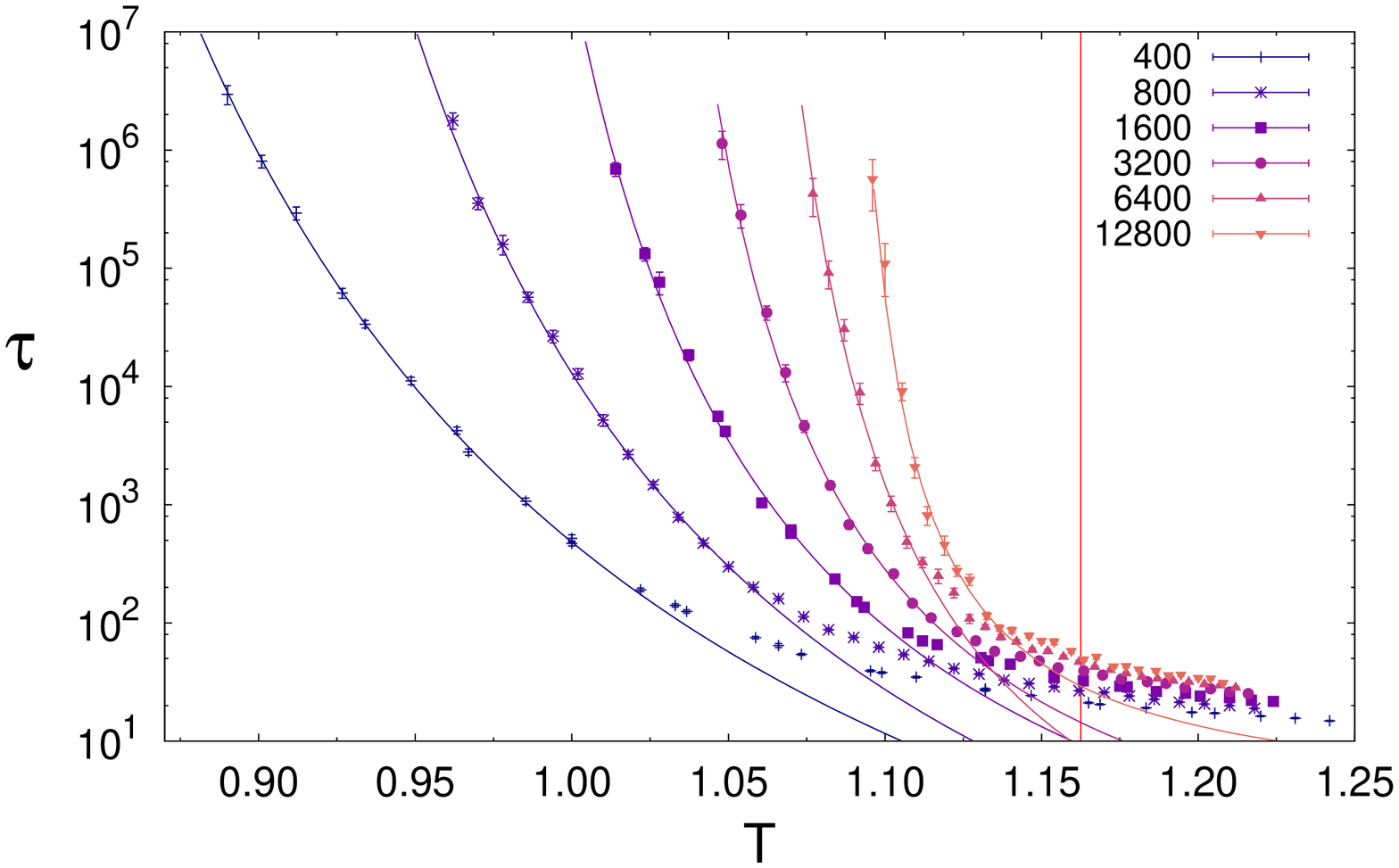}
\label{fig3a}
}
  \subfigure[]{
    \includegraphics[width=0.475\textwidth,height=0.32\textwidth]{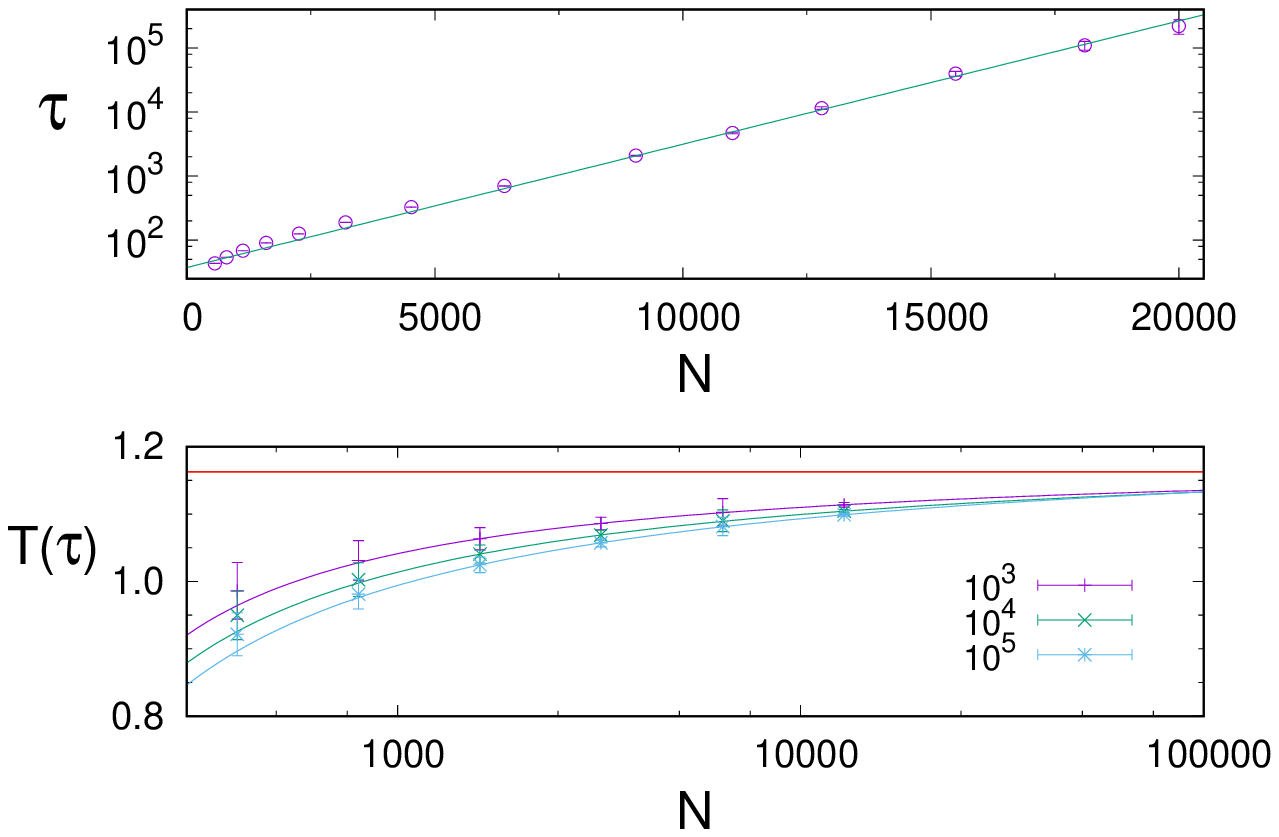}
    \label{fig3b}
  }
  \caption{ (a) Results obtained from Monte-Carlo simulations for the
    average time $\tau$ that it takes for the system to escape from a
    metastable anti-aligned state (see the main text) as a function
    of temperature. The values of the average connectivities are
    $c_\sigma=c_\tau=10$ and $c_I=1$, while the coupling strengths are
    given by $J_{\sigma} = J_{\tau} = U=1/c_{\rm eff}$, with $c_{\rm
      eff} = \frac{1}{3}(c_{\sigma} + c_{\tau} + c_I )$.  The
    different system sizes $N$ are indicated on the graph. Lines
    are fits to $\tau(T) = A \exp{\left(E/(T-T_0)\right)}$, and the
    vertical line corresponds to the theoretical result for the
    instability temperature $T_{\mathrm{ins}}=1.163$.  (b) The upper figure displays the behaviour of $\tau$
    as a function of $N$ for $T=1.105$, where the line corresponds to
    an exponential fit (see the main text). The lower figure shows the
    system size dependence of the temperatures for which each fit in
    figure (a) gives a fixed average time $\tau$. The curves are fits of the form
    $T(\tau)=T_0+a/\ln{(bN)}$, and the horizontal line is
    the instability temperature $T_{\mathrm{ins}}=1.163$ in the limit $N \rightarrow \infty$.}
\label{MC1}
\end{center}
\end{figure}

We have also computed the average or typical time $\tau$ for the
system to escape from a metastable anti-aligned state. By preparing
the system in an initial configuration corresponding to the zero
temperature anti-aligned solution, with $m_{\sigma}=1$ and $m_{\tau}=-1$, the parameter
$\tau$ counts the average number of MC steps that the system needs
to reach a configuration in which one of the magnetizations changes
sign. Averages are taken from 100 to 2000 different realizations of
the random graphs.
Figure \ref{fig3a} exhibits $\tau$ as a function of $T$ for
the same combination of model parameters as in figure \ref{ssd1}.
The simplest temperature dependence of these results is well described, especially in the region
where $\tau$ increases abruptly, by the Vogel-Fulcher law $\tau(T) = A \exp{\left(\frac{E}{T-T_0}\right)}$, where
the parameters $A$, $E$ and $T_0$ depend on $N$.

Figure \ref{fig3a} strongly indicates that, in the limit $N \rightarrow \infty$, the typical time $\tau$ diverges as
$T \rightarrow T_{\mathrm{ins}}^{+}$, where
$T_{\mathrm{ins}}$ is the temperature above which metastable states are absent in the thermodynamic limit. Such divergent behaviour is confirmed
in figure \ref{fig3b}, where we show the exponential divergence of $\tau=B\exp(bN)$ as a function of the system size for
$T = 1.105 <T_{\mathrm{ins}}$, with resulting fitting parameters $B=37(4)$ and $b=0.00044(1)$.
This typically characterizes a thermally activated process of crossing
free-energy barriers \cite{Barkema}, which is consistent with the mean-field character
of our model, or with the fact that the free-energy barriers separating different macroscopic states are proportional to $N$. Thus, as long
as $T <  T_{\mathrm{ins}}$ and $N \rightarrow \infty$, the system becomes trapped in the metastable states, once it is prepared
in a configuration close to them.

The instability temperature can be extracted from the simulation data by inverting the fits in figure \ref{fig3a} to obtain $T(\tau)$, the temperatures at which, for a given size, the system takes on average $\tau$ MC steps to cross the free-energy barrier. 
Data for $\tau=10^3$, $10^4$ and $10^5$ are shown in figure \ref{fig3b}. The extrapolation of $T_0$ for $N \rightarrow \infty$ can be performed by fitting the curves with $T_0 + a/\log{(bN)}$. The resulting values for $T_0$ are $1.19(1)$, $1.20(1)$ and $1.22(1)$, respectively, for $\tau=10^3$, $10^4$ and $10^5$, which agrees well with our theoretical result $T_{\mathrm{ins}}=1.163$, strictly valid for $N \rightarrow \infty$.

\section{Final remarks}

In this work we have studied the phase diagram and the existence of metastable states in a simple model
of coupled networks. The model is composed of two coupled Erd\"os-R\'enyi random graphs with an
Ising state variable or spin  lying at each node.
Each spin in a given graph has a finite number of
ferromagnetic couplings within its own graph and ferromagnetic interactions with a finite subset of spins located on the
other graph.
The simplicity of this model has enabled us to exactly compute the magnetization of each network and the free-energy
of the system in the thermodynamic limit, using the replica method of disordered systems. The main outcome
of our work is the full characterization of the phase diagram and of the stability properties of the different
macroscopic states.
As we clearly illustrate through the computation of the free-energy, the ferromagnetic
solution is the thermodynamic
state, while the anti-aligned solution (magnetizations with opposite signs) is always metastable.
The metastable solution appears
through a discontinuous transition as a function of the model parameters, provided the average connectivity
within each network is large enough and the temperature is sufficiently low.

We have estimated, through Monte-Carlo simulations, the average time $\tau$ the system needs
to escape from a metastable configuration. Our
results for $\tau(T)$ are well-described by the Vogel-Fulcher law, which tells us
that there is a critical temperature below which $\tau$ diverges
in the thermodynamic limit $N \rightarrow \infty$. Such ergodicity breaking stems
from the mean-field character of our model, in which the free-energy barrier between the metastable and
the stable macroscopic state diverges for $N \rightarrow \infty$.
Our main theoretical results have been compared
with Monte-Carlo simulations, showing a very good agreement.

We have also shown that the phase diagram, as
defined in the space of the connectivities of each network,
exhibits a low-temperature paramagnetic
phase for very small average connectivities.
This is explained by the fact that, in this sector of the phase diagram, the random graphs
are fragmented in a large number of disconnected finite clusters
that are unable to communicate. Finally, we point out that the present work paves the way to pursue a detailed
study of the cooperative behaviour arising in coupled networks with
different architectures \cite{Estrada,Peter2013}, such as modular and core-periphery structures. Work along
these lines is underway.

\ack

The authors thank Massimo Ostilli for a critical reading of the manuscript.
L.N. thanks the hospitality of Universidade Federal de Santa Maria, Brazil.
M.B. acknowledges a fellowship from CAPES.
F.L.M. thanks London Mathematical
Laboratory and CNPq (Edital Universal 406116/2016-4) for financial support. 

\section*{References}
\bibliographystyle{ieeetr} 
\bibliography{biblio}

\end{document}